\documentstyle[12pt]{article}
\newcommand{\be}{\begin{equation}}
\newcommand{\ee}{\end{equation}}
\newcommand{\ba}{\begin{eqnarray}}
\newcommand{\ea}{\end{eqnarray}}

\begin{document}
\begin{center}
{\large \bf  MULTIPARTICLE SUSY  QUANTUM MECHANICS AND THE REPRESENTATIONS OF
PERMUTATION GROUP.
}\\
\end{center}
\vspace{1cm}
M. V. Ioffe and A. I. Neelov\\ \\
{\small
Department of Theoretical
Physics, University of Sankt-Petersburg, 198904 Sankt-Petersburg, Russia.\\ \\
E-mail: ioffe@snoopy.phys.spbu.ru and neelov@snoopy.phys.spbu.ru\\ \\ \\

{\bf Abstract.\ \ }
The method of multidimensional SUSY Quantum Mechanics is applied
to the investigation of supersymmetrical
$N-$particle systems on a line for the case of separable center-of-mass
motion. New decomposition of the superhamiltonian into block-diagonal
form with elementary matrix components is constructed. Matrices of
coefficients
of these minimal blocks are shown to coincide with matrices of irreducible
representations of permutation group $S_N,$ which correspond to
the Young tableaux $(N-M, 1^M).$
The connections with known
generalizations of $ N-$particle Calogero and Sutherland models
are established.
}
\section*{\large\bf \quad 1. Introduction}
\hspace*{3ex}

One of the most natural generalizations of the standard \cite{witten}
1-dimensional Supersymmetrical Quantum
Mechanics (SUSY QM) concerns the systems in the spa\-ces of arbitrary
dimension $d$ \cite{abi}. It was shown for such systems that
superhamiltonian is a matrix $(2^d \times 2^d)$ block-diagonal operator
with $(d+1)$ components on the diagonal. These components of the
superhamiltonian are Schr\"odinger-type operators with matrix
$(C^n_d \times  C^n_d)$ potentials ($C^n_d -$ binomial coefficients,
$n=0,1,...,d$).
Supersymmetry of the system leads to important SUSY intertwining
relations between neighbouring components of the superhamiltonian
and provides definite connections between their spectra and eigenfunctions.
More definitely, for each component its spectrum consists of the
eigenvalues which coincide with a part of eigenvalues of neighbouring
components
of superhamiltonian. Corresponding eigenfunctions are connected with
each other by the action of supercharge operators (see details in
\cite{abi}). This approach was successfully used for some 2- and
3-dimensional physical systems \cite{appl}.

It would be interesting to apply this method for the systems with
another possible interpretation of several degrees of freedom in the
superhamiltonian. Namely, it seems to be useful in the description
of supersymmetrical systems of $N$ interacting quantum particles
on a line.

The supersymmetric generalization of a known exactly solvable
\cite{Calogero1}, \cite{Perelomov} N-particle Calogero model was
considered for the first time in the paper \cite{Freedman}, where its
spectrum was found (see also papers \cite{Efthimiou}, \cite{Ghosh},
\cite{Polychronacos}, \cite{Vasiliev1},
\cite{Vasiliev2}).
In the paper \cite{Efthimiou} the hypothesis was put forward
(but not proved),
that the (super)Calogero, Sutherland and some other models
possess shape-invariance \cite{gendenstein},
which could help us find the spectrum of the models
\footnote
{
However, some kind of shape-invariance for the
Calogero model was constructed in \cite{Ghosh} using the Dunkl operators
\cite{dunkl}.
}.

In the present paper the most general variety of supersymmetrical
$N-$particle systems on a line will be considered using the generalization
of method \cite{abi}. The only
restriction, introduced in Sect. 2, is the condition of separability
of center-of-mass
motion (CMM) \cite{Efthimiou} in the superpotential which seems to be
very natural for such systems.
Introducing usual bosonic Jacobi coordinates and their fermionic
analogues, we derive the superhamiltonian and SUSY intertwining
relations for systems with separable CMM. This block-diagonal
superhamiltonian has the same matrix dimension $2^N \times 2^N ,$
as in \cite{abi}, but more detailed structure: $2N$ blocks
$C^M_{N-1} \times C^M_{N-1}$ instead of $N+1$
blocks $C^M_N \times C^M_N$ in \cite{abi}.

In Sect. 3 the internal structure of the blocks on the diagonal
of the superhamiltonian is considered. It is shown that for any
$ M $ the coefficients
$B^{(M)}_{ij}$ in matrix potentials coincide with the matrices
of irreducible representation of permutation group $S_N,$ which
is characterized by the Young tableau $(N-M, 1^M).$  This statement
provides that these matrix potentials are elementary blocks of
the superhamiltonian, i.e. they cannot be further decomposed into
the block-diagonal form.
At the end of
Sect. 3 the SUSY intertwining relations are built in terms of Jacobi
coordinates, using the Clebsh-Gordan coefficients for the corresponding
irreducible representations of $S_N$.
Two examples are considered in Sect. 4. For the case
$ N=3 $ with particular choice of superpotential our approach gives a
part of the spectrum of $ 2\times 2 $ matrix Hamiltonian. The class
of superpotentials corresponding to $ N- $particle models with pairwise
interactions (including Calogero and Sutherland models) is
considered in the second example.
The connections with known
generalizations \cite{Freedman},\cite{Turbiner},\cite{Polychronacos},\cite{Vasiliev2} of $ N-$particle Calogero and Sutherland models
are established.
The proof of the Theorem of Sect. 3 can be found in Appendix.

\newpage
\section*{\large\bf 2.\quad Systems with a separable centre-of-mass motion }
\hspace*{3ex}
 The supersymmetric quantum system for arbitrary number of dimensions
$N$ consists \cite{abi}  of the superhamiltonian and the
supercharges\footnote { Here and below the indices $i,j,k,\ldots $
range from 1 to $N $.
}:
\ba
H_S = \frac{1}{2}(-\Delta + \sum_{i=1}^N(\partial_i W)^2-\Delta W)
+\sum_{i,j=1}^N\psi_i^+\psi_j\partial_i\partial_j W; \qquad \Delta\equiv
\sum_{i=1}^N \partial_i\partial_i;\ \ \partial_i\equiv \partial / \partial x_i;
\label{HSS}
\ea
\ba
Q^{\pm}\equiv
 \frac{1}{\sqrt{2}}\sum_{j=1}^N \psi_{j}^\pm (\pm\partial_j+\partial_j W); \label{lkq}
\ea
 with the algebra
\ba
H_S = \{Q^+,Q^-  \} ,\label{tri} \\
(Q^+)^2=(Q^-)^2=0, \label{che}
\ea
\be
[H_S,Q^\pm]=0, \label{sus}
\ee
  where $\psi_i,\ \psi_i^+$ are fermionic operators:
\be
\{\psi_i,\psi_j\}=0,\qquad
\{\psi_i^+,\psi_j^+\}=0, \qquad
\{\psi_i,\psi_j^+\}=\delta_{ij}.\label{ant}
\ee
The dynamics of a SUSY QM sys\-tem is de\-ter\-mined by a su\-per\-po\-ten\-tial
$ W$, depending on $N$ real coordinates $(x_1,\ldots,x_N)$.

For $N$-particle systems on a line it is natural to consider potentials with a separable centre-of-mass motion (CMM). Therefore in this paper we restrict ourselves by considering the superpotentials \footnote{The usefulness of the factor $1/\sqrt{N}$ will be explained later.}:
\be
 W(x_1,...,x_N)=w(x_1,\ldots,x_N)+W_C\biggl(\frac{x_1+...+x_N}{\sqrt{N}}\biggr)\ ;\quad
\sum_{j=1}^{N}\partial_j w(x_1,\ldots,x_N)=0, \label{sep}
\ee
allowing a separation of CMM motion ( $ w(x_1,\ldots,x_N)$ does not depend
\footnote{ The equation (\ref{sep}) is equivalent to the condition:
$
\sum_{k}\partial_k(\partial_i-\frac{1}{N}\sum_{l}\partial_l)w=0$ for every
$i=1,\ldots,N$. } on \\ $x_1+...+x_N$).

Let us introduce the operator:
\be
 \hat K_{ij}\equiv \psi^+_i\psi_j+\psi^+_j\psi_i-\psi^+_i\psi_i-\psi^+_j\psi_j+1=
1-(\psi_i^+-\psi_j^+)(\psi_i-\psi_j)= \hat K_{ji}=(\hat K_{ij})^\dagger, \nonumber
\ee
 which plays the role of the fermionic permutation operator
\ba
\hat K_{ij}\psi^+_i&=& \psi^+_j\hat K_{ij},\label{pe1} \\
\hat K_{ij}\psi^+_k&=&\psi^+_k \hat K_{ij},\quad k\ne i,j. \label{pr2}
\ea
In the fermionic Fock space
\be
 \psi_{i_1}^+\ldots\psi_{i_M}^+|0>\equiv |i_1\ldots i_M> ;\ \  \psi_{i}|0>=0
 \qquad i,i_1\ldots i_M ,M\leq N \label{bps}
\ee
this operator acts as:
\ba
\hat K_{ij}|i_1\ldots i \ldots j \ldots i_M>&=&|i_1\ldots j \ldots i \ldots i_M>\nonumber\\
\hat K_{ij}|i_1\ldots i \ldots i_M>&=&|i_1\ldots j \ldots i_M>\label{tij}\\
\hat K_{ij}|i_1 \ldots i_M>&=&|i_1 \ldots i_M>\qquad\mbox{\ for\ }\ \ i_1 \ldots i_M \neq i,j. \nonumber
\ea

Let us rewrite $H_S$ for the superpotentials (\ref{sep}) using $\hat K_{ij}$. We will take into account the
following equations:
\ba
\sum_{i,j=1}^{N}\psi_i^+\psi_j \partial_i \partial_jw=
\frac{1}{2}\sum_{i,j=1}^{N}\hat K_{ij}\partial_i\partial_jw=
 \frac{1}{2}\sum_{i\ne j}^N\hat K_{ij}\partial_i\partial_jw+
\frac{1}{2}\sum_{i=1}^N\partial_i^2w, \label{co1}
\ea
\ba
\sum_{i,j=1}^{N}\psi_i^+\psi_j
\partial_i \partial_j W_C=
\frac{1}{N}\biggl( \sum_{i=1}^N \psi_i^+\biggr) \biggl(  \sum_{j=1}^N \psi_j\biggr) W_C'',\label{co2}
\ea
\ba
\sum_{j=1}^N\biggl( \partial_j (w+W_C)\biggr)^2=
\sum_{j=1}^N
\biggl( \partial_j w \biggr)^2+(W_C')^2\label{co3}
\ea
to obtain
\ba
  H_S=
-\frac{1}{2}\Delta+\frac{1}{2} \sum_{j=1}^N
( \partial_j w)^2+
\frac{1}{2}\sum_{i\ne j}^N\hat K_{ij}\partial_i \partial_j w+ \nonumber \\
+\frac{1}{2}\bigl(
(W_C')^2-W_C''\bigr)+
\frac{1}{N} \biggl(\sum_{i=1}^N  \psi_i^+ \biggr)\biggl(\sum_{j=1}^N \psi_j \biggr) W_C''
.\label{HS0}
\ea

For the superpotentials (\ref{sep}) with a separable CMM  it is natural to go to the well-known Jacobi
coordinates\footnote{From this moment on, the variables denoted by letters $a,b,c,\ldots$ range from 1 to $N-1$.
} \cite{Reed}:
\ba
&y_b&=\frac{1}{\sqrt{b(b+1)}}(x_1+\ldots+x_b-bx_{b+1})
\label{bjc} \\
&y_N& = \frac{1}{\sqrt{N}}\sum_{i=1}^N x_i, \nonumber
\ea
or $y_k=\sum_{l=1}^NR_{kl}x_l$, where the matrix $R$ is determined by (\ref{bjc}).
The derivatives are connected by the same matrix: $\partial / \partial y_k=\sum_{l=1}^NR_{kl}\partial / \partial x_l, $ because $R$ is an orthogonal matrix.

For the supersymmetric systems it is natural to introduce also the
fermionic analogues of the Jacobi variables:


\ba
&\phi_b&=\frac{1}{\sqrt{b(b+1)}}(\psi_1+\ldots+\psi_b-b\psi_{b+1});
\nonumber\\
&\phi_N& = \frac{1}{\sqrt{N}}\sum_{i=1}^N \psi_i, \nonumber
\ea
or $ \phi_k=\sum_{l=1}^NR_{kl}\psi_l$, where the matrix $R$ is the same
as in (\ref{bjc}).
These variables also satisfy the standard anticommutation relations:
\be
\{\phi_k,\phi_l\}=0,\qquad
\{\phi_k^+,\phi_l^+\}=0, \qquad
\{\phi_k,\phi_l^+\}=\delta_{kl}.\nonumber
\ee

In terms of the Jacobi variables the supercharges (\ref{lkq}) can be rewritten as:
\ba
 Q^{\pm}= Q_C^{\pm}+ q^{\pm};
 \nonumber\\
Q_C^{\pm}\equiv
\frac{1}{\sqrt{2}}\phi^\pm_N\biggl(\pm\frac{\partial}{\partial
y_N}+W_C'\biggr); \qquad q^{\pm}\equiv
\frac{1}{\sqrt{2}}\sum_{b=1}^{N-1}\phi^\pm_b\biggl(\pm\frac{\partial}{\partial
y_b}+\frac{\partial}{\partial y_b}w\biggr). \label{145} \nonumber
\ea

Because
\be
\{ Q_C^{\pm}, q^{\pm}   \}=0, \label{ind}
\ee
the superhamiltonian $H_S$, acting in a $N$- dimensional superspace,
describes two non-interacting supersymmetric quantum systems:
\ba
  H_S= \{Q^+,Q^-  \} =
\{ q^+, q^-  \}+\{Q_C^{+}, Q_C^{-}\}
\equiv h+H_C,\label{HS1}
\ea
where
\ba
  h&=&\frac{1}{2}\sum_{b=1}^{N-1}\biggl(
-\frac{\partial^2}{\partial y_b^2}+
\biggl(\frac{\partial w}{\partial y_b}\biggr)^2-
\frac{\partial^2 w}{\partial y_b^2}\biggr)+
\sum_{b,c=1}^{N-1}\phi_b^+\phi_c
\frac {\partial^2 w}{\partial y_b \partial y_c},\label{ham}\\
  H_C&=& \frac{1}{2}\biggl(-\frac{\partial^2}{\partial y_N^2}+
(W_C')^2-W_C''\biggr)+\phi_N^+\phi_N W_C''    \label{H_N}
\ea
are  $(N-1)$- and $1$-dimensional SUSY Hamiltonians, respectively.
%

The Hamiltonian $h$ acting in the fermionic Fock space :
\be
  \phi_{b_1}^+\ldots\phi_{b_M}^+ |0>;\qquad b_i<b_j\ \mbox{for}\ i<j; \quad M<N,
\label{ste}
\ee
generated by fermionic creation operators $ \phi_b^+$, conserves
the cor\-res\-pon\-ding fer\-mionic num\-ber.
 Therefore, in the basis (\ref{ste}) it has \cite{abi} a
block-diagonal form:\\
$ h = diag(h^{(0)},..., h^{(N-1)})$, where matrix operator $ h^{(M)}$ of dimension $ C_{N-1}^{M}\times C_{N-1}^{M}$
is the component of $h$ in the subspace with fixed
fermionic number $M$.

In the same basis the supercharge $q^+$ changes \cite{abi} the fermionic number from $M$ to $M+1$ and has the following under-diagonal
structure:
\ba
 q^+=\pmatrix{
 0 & 0 & \ldots & 0 & 0 \cr
 q_{(0,1)}^+ & 0 & \ldots & 0 & 0  \cr
 0 & q_{(1,2)}^+  & \ldots & 0 & 0   \cr
 \vdots & \vdots & \ddots& \vdots & \vdots \cr
 0 & 0 & 0 & q_{(N-2,N-1)}^+  & 0  \cr}. \label{qpm}
\ea
 Similarly, $q^-=(q^{+})^{\dagger}$ is an over-diagonal matrix with nonzero elements\\
 $q_{(M+1,M)}^-=\biggl(q_{(M,M+1)}^+\biggr)^{\dagger}$.
Analogously, $H_C$ has a diagonal form
$H_C=diag(H_C^{(0)}, H_C^{(1)})$ in a basis
$(|0>,\phi_N^+|0>)$ and conserves the number of fermions $\phi_N$. In this case $H_C^{(0),(1)}$ are scalar (non-matrix) Hamiltonians. The one-dimensional supercharges $ Q_C^+, Q_C^-$ are a partial case of (\ref{qpm}) with off-diagonal  components $Q_{C(0,1)}^+, Q_{C(1,0)}^-$.
Superinvariance (\ref{sus}) of the superhamiltonian corresponds, in
components, to the intertwining relations \cite{abi} which can now be
decomposed as:
\ba
h q^+=q^+h  &\Leftrightarrow&
 h^{(M+1)} q_{(M,M+1)}^+= q_{(M,M+1)}^+ h^{(M)} \label{su1}\\
q^-h = h q^- &\Leftrightarrow&
q_{(M+1,M)}^- h^{(M+1)}= h^{(M)} q_{(M+1,M)}^- \label{su2}\\
H_C Q_C^+=Q_C^+H_C &\Leftrightarrow&
 H_C^{(1)} Q_{C(0,1)}^+=  Q_{C(0,1)}^+H_C^{(0)} \label{su3}\\
Q_C^-H_C = H_C Q_C^- &\Leftrightarrow&
 Q_{C(1,0)}^-H_C^{(1)}= H_C^{(0)} Q_{C(1,0)}^- \label{su4}
\ea
These intertwining relations lead \cite{abi} to the important connections
between spectra and eigenfunctions of "neighbouring" Hamiltonians, whose
fermionic numbers differ by 1. In particular, $ q_{(M,M+1)}^+\bigl(
q_{(M,M-1)}^-\bigr)$ maps eigenfunctions of $ h^{(M)}$ onto those of
$h^{(M+1)}( h^{(M-1)})$ with the same energy\footnote{ Let us note that the
eigenfunctions of $ h^{(M)}$ and $h^{(M+2)}$ are not connected directly by
supercharges $q^{\pm}$, contrary to the hypothesis of the paper
\cite{Efthimiou} in the context of Calogero-like models.  } (see details in
\cite{abi}).

In the total fermionic Fock space
\ba
 \phi_{b_1}^+\ldots\phi_{b_M}^+|0>, \quad
 \phi_{b_1}^+\ldots\phi_{b_M}^+\phi_N^+|0>;\ M < N \label{bph}
\ea
the superhamiltonian $H_S$ commutes with the operators $\sum_{b=1}^{N-1}\phi_b^+\phi_b $ and
$\phi_N^+\phi_N$ and therefore conserves the fermionic numbers of both
$\phi_b$ and $\phi_N$, separately.  Therefore, in this basis it has a
block-diagonal form, too:

\be
H_S=\pmatrix{
h ^{(0)}+H_C^{(0)}&\     &                      \ & \ & \ & \  \cr
                    \ &\ddots&                      \ & \ & \ & \  \cr
                    \ &\     &h ^{(N-1)}+H_C^{(0)}& \ & \ & \  \cr
\ & \ & \ &h ^{(0)}+H_C^{(1)}& \    & \                        \cr
\ & \ & \ &      \               &\ddots& \                        \cr
\ & \ & \ &                    \ & \    &h ^{(N-1)}+H_C^{(1)}  \cr }
\label{HSM}
\ee
where $ h ^{(M)},\ H_C^{(0),(1)}$ were determined above.

It is important that, due to (\ref{ind}), in the intertwining relations (\ref{su1})-(\ref{su4}) one can replace
$ h ^{(M)},\ H_C^{(0),(1)}$ by the components (\ref{HSM}) of $H_S$.

\section*{\large\bf 3.\quad  Internal structure of the components of the superhamiltonian }
\hspace*{3ex}

In the Eq. (\ref{HS0}) the superhamiltonian $H_S$ was written in the coordinates $(x_i,\psi_j)$ in terms of the fermionic permutation operator $\hat K_{ij}$.
In this section the structure of the blocks of $H_S$ (\ref{HSM})
in the basis (\ref{bph}) will be investigated.
In this basis the components of $H_S$ have the form:
\ba
  H^{(M)}_{S}=
-\frac{1}{2}\Delta+\frac{1}{2}\sum_{j=1}^N
\biggl( \frac{\partial w}{\partial x_j}\biggr)^2+
\frac{1}{2}\sum_{i\ne j}^N B_{ij}^{(M)}\partial_i\partial_j w+
\frac{1}{2}\bigl(
(W_C')^2\mp W_C''\bigr),\label{hm0}
\ea
where the matrices $ B_{ij}^{(M)}$  represent\footnote{
 It can be checked that $\hat K_{ij}$ conserves the fermionic numbers of both
$\phi_b $ and $\phi_N$, so it has the same block-diagonal structure in the basis (\ref{bph}) as $H_S$.}
the operator $\hat K_{ij}$ in the Fock subspace with fermionic number $M$.
Signs $\mp$ in (\ref{hm0}) correspond to the components
$(h ^{(M)}+H_C^{(0),(1)})$ of the superhamiltonian.
From this moment on we will consider the components of $H_S$ only in the form
(\ref{hm0}), i.e., in terms of the coordinates $(x_i,\phi_i^+)$. We prefer
$ \phi_i^+ $ to $\psi_i^+ $, because in terms of $ \phi_i^+ $ the block structure of $H_S$ is more detailed($2N$ blocks instead of $N+1$). The variables $x_i$ are preferable to $y_i$, because $x_i$ represent the coordinates of physical particles.
It is especially important in the cases when the particles are identical or the interaction is pairwise.

We move on to determining the matrices\footnote{Let us
stress that $ B_{ij}^{(M)}$ is not a matrix element, but a whole matrix of
dimension $ C^M_{N-1}\times C^M_{N-1}$.} $ B_{ij}^{(M)}$ .  The relations
(\ref{pe1}),(\ref{pr2}) can be rewritten as

\be
\hat K_{ij}\psi_k^+= \sum_{l} T_{(ij)lk}\psi_{l}^+\hat K_{ij},
\ee
 where
\be
T_{(ij)lk}\equiv
\delta_{lk}-\delta_{li}\delta_{ki}-\delta_{lj}\delta_{kj}
+\delta_{li}\delta_{kj}+\delta_{lj}\delta_{ki}.\label{per}
\ee
Applying this commutation rule to a state (\ref{bps}) $M$ times, we get:
\be
\hat K_{ij} \psi_{k_1}^+ \ldots\psi_{k_M}^+|0>=
\sum_{l_1,\ldots,l_M}T_{(ij)l_1k_1}\ldots T_{(ij)l_Mk_M} \psi_{l_1}^+\ldots\psi_{l_M}^+|0>.
\label{ten}
\ee
From the partial case of (\ref{ten}) with $M=1$ one can see that $T_{(ij)lk}$
is a matrix element of the permutation operator $\hat K_{ij}$ between one-fermionic states.

Thus, $\hat K_{ij}$ realizes a tensor representation  of rank $M$
of the symmetric
group $S_N$ of permutations of $\psi^+_i$
 on the states (\ref{bps}) with fixed fermionic number $M$. These states are obviously antisymmetric.

  Substituting  $\psi_l^+=\sum_{k}R_{kl}\phi_k^+$ into (\ref{ten}), one obtains:
\ba
  \hat K_{ij} \phi_{n_1}^+ \ldots\phi_{n_M}^+|0>=
\sum_{m_1,\ldots,m_M} \tilde T_{(ij)m_1n_1}\ldots \tilde T_{(ij)m_Mn_M}
\phi_{m_1}^+ \ldots\phi_{m_M}^+|0>,\label{tp0}
\ea
where \footnote
{It should be stressed that $\hat K_{ij}$ in
(\ref{tp0}) permutes $\psi_{k}^+$, not $\phi_{k}^+$.}
\be
  \tilde T_{(ij)mn}\equiv \sum_{k,l}R_{mk} T_{(ij)kl} R_{nl}. \label{til}
\ee
Note that $\hat K_{ij}\phi^+_N= \phi^+_N\hat K_{ij}$, so it is enough to
consider $\hat K_{ij}$ in the subspace
\be
\phi_{b_1}^+\ldots\phi_{b_M}^+|0>;\ M < N . \label{st1}
\ee
It is therefore possible to rewrite (\ref{tp0}) as
\ba
  \hat K_{ij} \phi_{a_1}^+ \ldots\phi_{a_M}^+|0>=
 \sum_{b_1,\ldots,b_M}\tilde T_{(ij)b_1a_1}\ldots \tilde T_{(ij)b_Ma_M}
\phi_{b_1}^+ \ldots\phi_{b_M}^+|0>.\label{tph}
\ea

Thus, in the space spanned by the states (\ref{st1}) the operators $K_{ij}$
and matrices $ B_{ij}^{(M)}$ also realize a tensor representation of rank $M$
of the symmetric
group $S_N$ of permutations of $\psi^+_i$ (the states (\ref{st1}) are also antisymmetric).

 In the Appendix we prove by induction the

{\it \bf Theorem}:
the representation (\ref{tph}) of the group $S_N$ of permutations of $
\psi^+_i $ is irreducible and corresponds to the Young tableau\footnote {
The standard notation \cite{Hammer} for a Young tableau containing
$\lambda_i$ cells in the $i$-th line is $(\lambda_1,\ldots,\lambda_n)$; if
the tableau contains $m$ identical lines with $\mu$ cells, it is denoted
by $(\ldots,\mu^m,\ldots)$.  } $(N-M,1,\ldots,1)\equiv (N-M,1^M)$.

It is clear that the $2N$ blocks $(h ^{(M)}+H_C^{(0),(1)})$ in (\ref{HSM})
are "subblocks" of the $N+1$ components of $H_S$ that would be obtained if
we developed a standard supersymmetric formalism \cite{abi} for $H_S$ in
the coordinates $x_i,\ \psi^+_i$. The natural question is whether there
exist even smaller "subblocks", or the blocks (\ref{hm0}) are "elementary".
Because for any $M$, due to the statement of the Theorem, the matrices $
B_{ij}^{(M)}$ realize an irreducible representation of $S_N$, they cannot
be simultaneously made block-diagonal by a change of basis. In general
case, when all their coefficients $\partial_i\partial_j w$ are independent,
it means that $H^{(M)}_{S}$ cannot in turn be made block-diagonal by any
change of basis.  Thus, the blocks (\ref{hm0}) are "elementary".

According to (\ref{su1})-(\ref{su4}), these elementary blocks of $H_S$ are intertwined by the components
$ q_{(M,M+1)}^+,q_{(M,M+1)}^-,Q^+_{(0,1)}, Q^-_{(1,0)}$ of supercharges.
Let us now consider the properties of $ q_{(M,M+1)}^+,q_{(M,M+1)}^-$ in the context of permutation group $S_N$.

 Being the components of
\be
 q^{+}=\sum_{b=1}^{N-1}\phi_b^{+} a_b^{+};\qquad a^{+}_b\equiv \frac{1}{\sqrt{2}}(
\partial/ \partial y_b + \partial w/ \partial y_b), \label{ori}
\ee
the operators $q_{(M,M+1)}^+$ map the  eigenfunctions $\Psi^{(M)}$ of the
subspace with fermionic number $M$ into the  eigenfunctions $\Psi^{(M+1)}$
of the subspace with fermionic number $M+1$. Let $\Psi^{(M)}_\nu$ be the
components of $\Psi^{(M)}$ in some basis of the subspace (\ref{st1}). The
exact form of the basis is unimportant, though it is necessary that the
matrices $B_{ij}^{(M)}$ and the Clebsh-Gordan coefficients introduced
below be defined in this basis too. Because the dimension of the subspace
(\ref{st1}) is $C_{N-1}^M$, $\nu$ ranges from 1 to $C_{N-1}^M$.

Then the operator $ q_{(M,M+1)}^+$ takes
a  matrix form:

\be
\Psi^{(M+1)}_\mu= (q_{(M,M+1)}^+\Psi^{(M)})_\mu=
\sum_{\nu=1}^{C_{N-1}^M} (q_{(M,M+1)}^+)_{\mu\nu}\Psi^{(M)}_\nu. \nonumber
\ee
In the same way,
\be
(\phi_b^+\Psi^{(M)})_\mu=\sum_{\nu=1}^{C_{N-1}^M} ( \phi_b^+)_{\mu\nu}\Psi^{(M)}_\nu.
\nonumber
\ee
Therefore (\ref{ori}) can be rewritten as
\be
\biggl(q_{(M,M+1)}^+\biggr)_{\mu\nu}= \sum_{b=1}^{N-1}a^{+}_{b} ( \phi_b^+)_{\mu\nu}.\nonumber
\ee

We know that $\phi_b^+$
is transformed under permutations of $\psi_i^+ $ as an irreducible representation of $S_N$ with a Young tableau
$(N-1,1)$.Therefore, the matrix element
$( \phi_b^+)_{\mu\nu}= (M\ \mu,1\ b|M+1\ \nu)$, where $ (M\ \nu,1\ b|M+1\ \mu )$ are Clebsh-Gordan coefficients which correspond to the transition\footnote{$\times$ denotes the interior product of Young tableaux.}:
$(N-M,1^M)\times(N-1,1)\rightarrow (N-M-1,1^{M+1})$.
In the notations of \cite{Hammer} the symbols $M,M+1,1$ in the Clebsh-Gordan coefficients correspond to the representations
of $S_N$ with the Young tableaux
${(N-M,1^M),(N-M-1,1^{M+1}),(N-1,1)}$, respectively.
Finally,
\be
\biggl(q_{(M,M+1)}^+\biggr)_{\mu \nu }= \sum_{b=1}^{N-1} a^{+}_{b}\cdot (M\ \nu ,1\ b|M+1\ \mu ).\label{qa1}
 \ee
Similarly,
\be
\biggl(q_{(M+1,M)}^-\biggr)_{\rho \sigma }=\sum_{b=1}^{N-1} a^{-}_{b}\cdot (M+1\ \sigma ,1\ b|M\ \rho );\qquad a_b^-=(a_b^+)^{\dagger}\label{qa2}
 \ee
In the $x_i$ coordinates, $ a^{\pm}_{b}=\sum_{l=1}^NR_{bl} A_l^{\pm}$ where
$ A_l^{\pm}\equiv \frac{1}{\sqrt{2}}(\pm\partial_l+\partial_l w)$.

Substituting $H_S^{(M)}$ and (\ref{qa1})-(\ref{qa2}) into the intertwining relations (\ref{su1})-(\ref{su2}), one obtains:
\ba
 \sum_{b=1}^{N-1}\sum_{\mu=1}^{C_{N-1}^{M+1}}\sum_{l=1}^N H^{(M+1)}_{\rho \mu } R_{bl} A_l^{+} (M\ \sigma ,1\ b|M+1\ \mu )= \nonumber\\
=\sum_{c=1}^{N-1}\sum_{\nu=1}^{C_{N-1}^{M}}\sum_{m=1}^N R_{cm} A_m^{+} (M\ \nu ,1\ c|M+1\ \rho )H^{(M)}_{\nu \sigma }; \label{sj1}\\
\sum_{b=1}^{N-1}\sum_{\sigma=1}^{C_{N-1}^{M+1}}\sum_{l=1}^N R_{bl} A_l^{-} (M+1\ \sigma ,1\ b|M\ \rho) H^{(M+1)}_{\sigma \epsilon }=\nonumber\\
=\sum_{c=1}^{N-1}\sum_{\delta=1}^{C_{N-1}^{M}}\sum_{m=1}^N H^{(M)}_{\rho \delta } R_{cm} A_m^{-} (M+1\ \epsilon ,1\ c|M\ \delta ). \label{sj2}
\ea

Now it follows from (\ref{sj1})-(\ref{sj2}) that the eigenfunctions $\Psi^{(M)}$ and $\Psi^{(M+1)}$ are connected by
\be
\Psi^{(M+1)}_\mu = \sum_{b=1}^{N-1}\sum_{l=1}^N \sum_{\nu=1}^{C_{N-1}^{M}} R_{bl} A_l^{+} (M\ \nu ,1\ b|M+1\ \mu ) \Psi^{(M)}_\nu;\label{rel}
\ee
\be
\Psi^{(M)}_\rho = \sum_{c=1}^{N-1}\sum_{m=1}^N \sum_{\sigma=1}^{C_{N-1}^{M+1}} R_{cm} A_m^{-} (M+1\ \sigma ,1\ c|M\ \rho ) \Psi^{(M+1)}_\sigma.\nonumber
\ee
The relations (\ref{su3}),(\ref{su4}) remain unchanged, because the supercharges
 $ Q^+_{(0,1)}, Q^-_{(1,0)} $ are scalar operators.

\section*{\large\bf 4.\quad Examples.}
\subsection*{\normalsize 4.1.\quad 3-particle supersymmetrical system }
\vspace*{1ex}
In the case of $N=2$ the superhamiltonian $h$ corresponds to the standard one-dimensional SUSY QM.
For the simplest
 non-trivial case of $ N = 3 $ the superhamiltonian
(\ref{HSM}) consists of six components: four scalar and two $ 2 \times 2 $
matrix Schr\"odinger-type operators. Their spectra and eigenfunctions are
connected with each other due to SUSY intertwining relations.

Let us consider a simple system, generated by the superpotential
\be
 W =- \ln \biggl(3+a \sum_{j<k}^3(x_j-x_k)^2\biggr)-
\frac{ a}{2}\sum_{i=1}^{3}x_i^2; \quad a > 0 . \label{chi}
\ee
This superpotential allows the separability of CMM (\ref{sep}) with
\be
W_C =- \frac {a}{2}y_3^2;\nonumber
\ee
\be
w =- \ln (3+a \sum_{j<k}^3(x_j-x_k)^2)-
\frac{ a}{6}\sum_{j<k}^{3}(x_j-x_k)^2 \label{ch<}
\ee
Two of scalar blocks (\ref{hm0}) of the corresponding superhamiltonian
will then take the form of 3-dimensional harmonic oscillator with well-known spectrum:
\ba
 H_{S}^{(0)}= \frac{1}{2}\biggl( \sum_{i=1}^{3}\partial_i^2+
a^2\sum_{i=1}^{3}x_i^2\biggr)
\pm\frac{a}{2},\label{eh0}
\ea
Two other scalar components of (\ref{HSM}) are not exactly solvable:
\be
H^{(2)}_S=\frac{1}{2}\biggl(-\Delta+
a^2\sum_{i=1}^{3}x_i^2\biggr) -
\frac{36a}{(3+a \sum_{j<k}^3(x_j-x_k)^2)^2}\pm \frac{a}{2}.\label{eh2}
\ee
Nevertheless, SUSY intertwining relations (\ref{sj1}) - (\ref{sj2}) give
us the opportunity
to find the part\footnote{In some sense, this situation resembles
so called quasi-exactly-solvable models \cite{turbiner}, for which only a part of eigenstates and eigenfunctions is known.} of
spectrum
of both matrix components (\ref{hm0}):
\be
H^{(1)}_S=\frac{1}{2}\biggl(-\Delta+
a^2\sum_{i=1}^{3}x_i^2\biggr) -
\frac{18a+12a^2\sum_{i<j}^3 B_{ij}^{(1)}(x_i-x_j)^2}
{ (3+a \sum_{j<k}^3(x_j-x_k)^2)^2}\pm\frac{a}{2},\label{eh1}
\ee
where matrices $ B_{ij}^{(1)}$ realize a simple irreducible
representation of group $ S_{3}:$
\ba
B^{(1)}_{12}=\pmatrix{
 -1  &       0                  \cr
 0  &      1                   \cr };\qquad
B^{(1)}_{23}=\pmatrix{
 1/2  &    \sqrt{3}/2                  \cr
  \sqrt{3}/2  &      -1/2                   \cr };\qquad
B^{(1)}_{13}=\pmatrix{
 1/2  &    -\sqrt{3}/2                  \cr
  -\sqrt{3}/2  &      -1/2                   \cr }.
\nonumber
\ea
These Hamiltonians (\ref{eh0})-(\ref{eh2}) are intertwined
(see (\ref{sj1}), (\ref{sj2})) by
the supercharges (\ref{qa1}),(\ref{qa2}), where
\be
A_l^{\pm}=\frac{1}{\sqrt{2}}\biggl(\pm\partial_l-
\frac{2a(3x_l-\sqrt{3}y_3)}{3+a \sum_{j<k}^3
(x_j-x_k)^2}-a(x_l-\sqrt{3}/3y_3)\biggr).
\nonumber
\ee
The Clebsh-Gordan coefficients can be found, for example, in
\cite{Hammer}. In the case of $N=3$ one can write them
explicitly:
\ba
(0\ \nu ,1\ b|1\ \mu )= (1\ \mu ,1\ b|0\ \nu )=
\delta^{1}_{\mu}\delta^{2}_{b}+\delta^{2}_{\mu}\delta^{1}_{b}\ \mbox{where\
\ }\nu= 1; \nonumber\\
(2\ \sigma ,1\ b|1\ \rho )=
 (1\ \rho ,1\ b|2\ \sigma )= \delta^{1}_{\rho}\delta^{2}_{b}-\delta^{2}_{\rho}\delta^{1}_{b}\ \mbox{where\
\ }\sigma= 1.\label{mis}
\ea
In these expressions $\nu=1$ and $\sigma=1$  means that these indices span
the basis of the  representations of $S_3$, corresponding to
the Young tableaux $(3)$ and $(1^3)$, respectively. These
representations are one-dimensional.

The generalization onto higher $N$ is straightforward. Let us note, that
the above approach can also be applied without any change to the systems, which are not
symmetric under permutations of $x_i$.

\subsection*{\normalsize 4.2.\quad $ N $-particle supersymmetric
systems with a pairwise interaction.}
\vspace*{1ex}
If we are interested in the scalar and matrix
Hamiltonians $H_{S}^{(M)}$ with a pairwise interaction\footnote
{
We restrict ourselves here by considering
the superpotentials $w$ that are symmetric
under permutations of $x_i$, though this approach admits direct
generalization to non-symmetric superpotentials with separable CMM.
}, it is necessary (but insufficient) to consider
superpotentials such that \footnote{We imply here that
$ W_{C} = 0 $, except for the well-known Calogero model (see below).}
\be
\partial_i\partial_j w=f(x_i-x_j);\ i\ne j,\label{str}
\ee
where $f(x)$ is some real function. It means that
\be
w=\sum_{i< k}^NU(x_i-x_j)+\sum_{j=1}^N h(x_j), \nonumber
\ee
where $U(x), h(x)$ are also real functions.

We will restrict ourselves futher by considering only the case of
\be
 w=\sum_{i < j}^NU(x_i-x_j);\qquad U(x)=U(-x).   \label{pss}
\ee

For such $w$ components $H_{S}^{(M)}$ (see (\ref{hm0})) have the form

\ba
  H^{(M)}_{S}=
\frac{1}{2}\biggl[-\Delta+ \sum_{i\ne l}^N(U_{il}')^2+
\sum_{i\ne l_1 \ne l_2 \ne i}^N U_{il_1}' U_{il_2}'-
\sum_{i\ne j}^N B_{ij}^{(M)}U_{ij}''\biggr],\label{hmp}
\ea
where $U_{ij}\equiv U(x_i-x_j)$.

These matrix Hamiltonians are intertwined by relations (\ref{sj1}),
(\ref{sj2}), where
\be
A_l^{\pm}=\frac{1}{\sqrt{2}}(\pm\partial_l+\sum_{k \neq l}^NU'(x_l-x_k)).
\nonumber
\ee

 For $ H^{M}_{S}$ to be pairwise, it is also necessary
\cite{Perelomov},\cite{Efthimiou},
that $\sum_{i\ne l_1 \ne l_2 \ne i}^N U_{il_1}' U_{il_2}'$
decompose into a sum of pairwise terms. Therefore
there should exist a real function  $v_0(x):$
\be
  U'(A)U'(B)+U'(A)U'(C)+U'(B)U'(C)=v_0(A)+v_0(B)+v_0(C). \label{eqp}
\ee
Then the scalar term in (\ref{hmp}) has the form:
$$\sum_{i\ne l}^N\biggl[(U'(x_i-x_l))^2+v_0(x_i-x_l)\biggr]\equiv
 \sum_{i\ne l}^N V(x_i-x_l).$$

All solutions of (\ref{eqp}) were found by Calogero \cite{Calogero2}. They are given in the Table (See \cite{Efthimiou}).
\begin{table}
{Table}
\begin{tabular}{ccccc}
\hline
\hline
$ U(x)$ &\ \ \ \ \ \ &$V(x)$ &\ \ \ \ \ \ &$U''(x)$ \\
\\
\hline
$\frac{1}{2}ax^2+b\ln(x)$
&\ \ \ \ \ \ \ \ \ \ \ \ \  \ &$\frac{b(b+1)}{x^2}+\frac{3}{2}a^2x^2 + 3ab+a$ &\ \ \ \ \ \ \ \ \ \ \ \ \ \ \ \ \ \ &$a-\frac{b}{x^2}$ \\
\\
 $a|x|$ &\ \ \ \ \ \ \ \ \ &$a^2sgn^2x+a\delta(x)-\frac{1}{3}a^2$ & \ \ \ \ \ \ \ \ & $2a\delta(x)$\\ \\
$a\ln\sin x$ &\ & $\frac{a(a-1)}{\sin^2x}-\frac{4}{3}a^2$ &\ &
 $-\frac{a}{\sin^2x}$ \\ \\
$a\ln \sinh x $&\ &$ \frac{a(a-1)}{\sinh^2 x}+ \frac{2}{3}a^2$&\ &
$-\frac{a}{\sinh^2x}$ \\  \\
$\ln |\theta_1(\frac{\pi x}{2\omega}|\frac{ir}{2\omega})|$&\ &
$(a+\frac{1}{2})P(x)+\frac{\zeta(\omega)}{\omega}$ &\ &
$aP(x)-\frac{\zeta(\omega)}{\omega}$\\ \\
\hline
\hline
\end{tabular}
$\zeta(x)$ is the Weierstrass $\zeta$ function;
$\zeta'(x)=-P(x)$ and $\omega$ is half-period \cite{Perelomov}.
\end{table}
So, the components (\ref{hmp}) are now pairwise and take the form:
\ba
  H^{(M)}_{S}=
\frac{1}{2}\biggl[-\Delta+ \sum_{i\ne l}^N V(x_i-x_l)-
\sum_{i\ne j}^N B_{ij}^{(M)}U_{ij}''\biggr]. \label{typ}
\ea
Let us note, that to obtain standard Calogero model \cite{Calogero1}
$ (U(x)=ax^{2}/2+b\ln x), $ we have to add into (\ref{typ}) the terms:
$$ \frac{1}{2}(W_{C}^{\prime 2}\mp W_{C}^{\prime\prime})=\frac{1}{2}(a^{2}N^{2}y_{N}^{2}
\mp aN), $$ corresponding to nonzero $W_C(y_N)=\frac{1}{2}aNy_N^2$.

In the cases of superpotentials  from the Table, which correspond
to Calogero and Sutherland models,
spectrum of the superhamiltonian $ H_{S} $ was obtained in
\cite{Freedman}, \cite{Turbiner}.
For the same superpotentials,
components (\ref{typ}) of the superhamiltonian coincide with
(also exactly solvable)
matrix generalizations of Calogero and Sutherland models
\cite{Polychronacos},\cite{Vasiliev2} in partial case of
the Young tableaux $ (N-M, 1^{M})$. These tableaux were obtained
in the Theorem in the present paper (see Appendix).

\newpage

\section*{\large\bf 5.\quad Appendix}
\hspace*{3ex}
In this Appendix, we will prove the following
\vspace{1ex}

{\it \bf Theorem}: the operator $\hat K_{ij}$ realizes on the states
 (\ref{st1})
 the irreducible representation of the group $S_N$ of permutations of $\psi^+_i$, corresponding to the Young tableau $(N-M,1^M)$.
\\
At first, it is necessary to prove the Lemma, corresponding to the partial case of the Theorem
for $M=1$:
\vspace*{1ex}

{\it \bf Lemma}:\ \ the operator $\hat K_{ij}$ (and matrices $\tilde T_{(ij)}$ (\ref{til})), acting on the states $\phi_{b}^+|0>$, realize  the irreducible representation of the group $S_N$ of permutations of $\psi^+_i$, corresponding to the Young tableau  $(N-1,1)$.

  Proof: by induction on $N$.

  For $N=2$ we have only one Jacobi coordinate:
$\phi_1^+=\frac{1}{\sqrt{2}}(\psi_1^+-\psi_2^+)$. Obviously, the state
$\phi_1^+|0> $ is transformed as a representation of $S_2$, with
a Young tableau $(1^2)$.

  Let us suppose  that $\phi_{1}^+|0>, \ldots,\phi_{N-2}^+|0>$ form a
representation of the group $S_{N-1}$ of permutations\footnote
{Here and below, the (physical) coordinates $x_{i'},\psi_{i'}$ of the first $N-1$ particles are
denoted by indices $i',j',k',\ldots $. Let us stress that they are not the
Jacobi coordinates $y_a,\phi_a$.} of $\psi^+_{i'};\ i'<N$,
corresponding to a Young tableau $(N-2,1)$.  It is to be proved  that
$\phi_{1}^+|0>, \ldots,\phi_{N-1}^+|0>$ form a representation of the group
$S_N$ of permutations of $\psi^+_i$, corresponding to a Young tableau
$(N-1,1)$.

     (i) We follow the method from the book \cite{Hammer} of construction
of the irreducible representations of $S_N$ when the
representations of $S_{N-1}$ are known\footnote
{
    In \cite{Hammer} another numeration of basis vectors was used.
}.
It uses the fact that arbitrary permutation
$ \hat K_{ij}$ of $ \psi^+_i, \psi^+_j $ is a combination
of the permutations $\hat K_{i'j'}$ where $i',j'<N$ and $\hat K_{N-1,N} $. The only
nontrivial case here is

 \be
 \hat K_{j'N}=\hat K_{N-1,N}\hat K_{j',N-1} \hat K_{N-1,N}.
 \nonumber
 \ee

The method of \cite{Hammer}, applied to a representation with a Young
tableau $(N-1,1)$(we do not reproduce here the long proof) yields the
following result:  the subgroup $S_{N-1}$ of $S_N$ consisting of the
permutations $\hat K_{i'j'}$ is realized in this representation by
$(N-1)\times(N-1)$ matrices
\be
\pmatrix{ U_{(i'j')} & 0  \cr
           0     & 1  \cr },\nonumber
\ee
where $U_{(i'j')}$ are the matrices of representation of $S_{N-1}$
 corresponding to a Young tableau $(N-2,1)$; their components are
$U_{(i'j')ab}$, but here and below we will not write the indices $a,b$ for
brevity.  The permutation $\hat K_{N-1,N}$ is realized in this
representation by the $(N-1)\times(N-1)$ matrix
\be
\pmatrix{  1 & \ & \ & \ & \   \cr
           \ & \ddots & \ & \ & \   \cr
           \ & \ & 1 & \ & \   \cr
           \ & \ & \ & \frac{1}{N-1} & \frac{\sqrt{N(N-2)}}{N-1}   \cr
           \ & \ & \ & \frac{\sqrt{N(N-2)}}{N-1} & -\frac{1}{N-1}   \cr }.\nonumber
\ee

       (ii) It is easy to show that
in the terms of the Jacobi variables the permutations of
 $\psi_{i'}^+$ are realized
by the following matrices:
\be
\pmatrix{ \tilde T_{(i'j')}^{(N-1)} & 0  \cr
           0     & 1  \cr },\nonumber
\ee
where $ \tilde T^{(N-1)}_{(i'j')}$ is equal to $\tilde T_{(ij)}$ from the
previous step of induction. Because of the assumption of induction, $\tilde
T^{(N-1)}_{(i'j')}=U_{(i'j')}$.

The permutation $\hat K_{N-1,N}$ is realized (see (\ref{til})) by the following matrix:
$ \tilde T_{(N-1,N)bc}\equiv \sum_{j,l=1}^{N}R_{bj} T_{(N-1,N)jl} R_{cl}$. Substituting
$ T_{(N-1,N)jl} $ from (\ref{per}) ($R$ is the same as above) we obtain:
\be
   \tilde T_{(N-1,N)}=
\pmatrix{  1 & \ & \ & \ & \   \cr
           \ & \ddots & \ & \ & \   \cr
           \ & \ & 1 & \ & \   \cr
           \ & \ & \ & \frac{1}{N-1} & \frac{\sqrt{N(N-2)}}{N-1}   \cr
           \ & \ & \ & \frac{\sqrt{N(N-2)}}{N-1} & -\frac{1}{N-1}.   \cr }\nonumber
\ee
Permutation $\hat K_{bN}$ can be decomposed into a combination of these.
Thus, the proof of the Lemma is completed.
\vspace{1ex}

Now we can prove the Theorem by induction on $M$.

 For $M=1$ the statement of the Theorem is satisfied due to Lemma.

Let us suppose that, for some fixed $M$, on the states $ \phi_{b_1}^+ \ldots\phi_{b_M}^+|0>$ the irreducible representation of
the group $S_N$ of permutations of $\psi^+_i$ with $ i \leq N$, corresponding to
the Young tableau $(N-M,1^M),$ is realized.

Then we use this assumption to prove that on  the states $ \phi_{b_1}^+ \ldots\phi_{b_{M+1}}^+|0>$  the irreducible representation of the group $S_N$, corresponding to the Young tableau $(N-M-1,1^{M+1})$, is realized .

(i) As we have mentioned above, (\ref{st1}) is transformed under permutations from $S_N$ as an antisymmetric tensor. Such tensor can be provided with a Young tableau, describing the symmetry of its indices, namely,
$[1^M]$(we deliberately use different brackets\footnote
{
Let us stress that every state (\ref{st1}) is described by two principally
different Young tableaux: $[1^M]$, denoting its index structure as
antisymmetric tensor, and $(\lambda_1,\ldots,\lambda_N)$, describing it as
an irreducible representation of the symmetric group $S_N$ (the
notations are taken from \cite{Hammer}). }).

The assumption of induction affirms that (\ref{st1}) transforms under $S_N$
as an irreducible representation with a Young tableau $(N-M,1^M)$.

Let us consider a tensor product of the states (\ref{st1}) and
$\phi_{b_{M+1}}^+|0>$ and show that it contains
$\phi_{b_1}^+ \ldots\phi_{b_{M+1}}^+|0>$.

The tensor product will contain the tensors whose index structure
corresponds to the Young tableaux, contained in the so-called exterior
product \cite{Hammer} of Young tableaux, corresponding to the index
structure of the states (\ref{st1}) and $\phi_{b_{M+1}}^+|0>$ ,
namely\footnote{$\otimes,\,\times$ denote the exterior and interior products of Young tableaux, respectively.},
$[1^M]\otimes [1]$.

If  some Young tableau $[D]$ is contained in $[1^M]\otimes [1]$,
then, the corresponding tensor representation of $S_N$ is contained
in the tensor product of the tensor representation of $S_N$
corresponding to (\ref{st1}) and one corresponding to $\phi_{b_{M+1}}^+|0>$.
But these representations can also be considered as
irreducible representations of $S_N$ with Young tableaux $(N-M,1^M)$ and
$(N-1,1)$. The tensor representations from the tensor product may also
be decomposed into a direct sum of irreducible representations of
$S_N$ corresponding to some Young tableaux. The last Young tableaux
form a so-called
interior product:
 $(N-M,1^M)\times(N-1,1)$.

So the Young tableaux obtained after decomposition of a tensor
corresponding to $[D]$ are contained in $(N-M,1^M)\times(N-1,1)$.

(ii) It is shown in \cite{Hammer} that for any Young tableau
$[\lambda_1\ldots \lambda_k]$
\be
 [\lambda_1\ldots \lambda_k]\otimes [1]=
\sum_{i=1}^{k}[\lambda_1\ldots,\lambda_{i-1},\lambda_i+1,\lambda_{i+1}\ldots \lambda_k]. \label{pro}
\ee
So, for the state (\ref{st1}) we get:
\be
   [1^M]\otimes[1]=[1^{M+1}]\oplus[2,1^{M-1}].\label{pr1}
\ee
The state $\phi_{b_1}^+ \ldots\phi_{b_{M+1}}^+ |0>$ has an index  structure
described by the Young tableau $[1^{M+1}]$ and is a direct sum of
irreducible representations of $S_N$ corresponding to some Young tableaux.
We will further denote the set of these tableaux as (A). Taking into
account the considerations from the point (ii), we can state that $
(N-M,1^M)\times(N-1,1)=(A)\oplus\ldots$.

(iii) It is known \cite{Hammer} that the interior product of every Young
tableau with $(N-1,1)$ contains only the Young tableaux differing from the
initial one by no more than the position of one cell. (Let us remind that
all the Young tableaux describing the irreducible representations of $S_N$
contain exactly $N$ cells each.) Therefore,\\ $(N-M,1^M)\times(N-1,1)$ may
contain only the following Young tableaux:  $ (N-M,1^M),$
$(N-(M+1),1^{M+1}),$ $(N-M,2,1^{M-2}) ,$ $(N-(M+1),2,1^{M-1})$.

(iv)It is useful here to consider the state
$\phi_{b_1}^+ \ldots\phi_{b_{N-1}}^+|0>$. Its index structure corresponds to a Young tableau $[1^{N-1}]$. As a representation of $S_N$, this state corresponds to
a Young tableau $(1^N)$,because
\ba
  \phi_{1}^+ \ldots\phi_{N-1}^+|0>=
 \phi_{1}^+ \ldots\phi_{N-1}^+\phi_{N}\phi_{N}^+|0>
=(-1)^{N-1} \phi_{N}\phi_{1}^+ \ldots \phi_{N}^+|0>=\nonumber\\
=C \phi_{N}\psi_{1}^+ \ldots\ \psi_{N}^+|0>,\label{NNN}
\ea
where $C$ is some nonzero constant. It is obvious that $\hat K_{ij}$, acting upon (\ref{NNN}), changes its sign.

(v) Let us consider a tensor product of $\phi_{b_1}^+ \ldots\phi_{b_M}^+|0> $
with $\phi_{b_{M+1}}^+|0> $, then once more a tensor product of the result
with $\phi_{b_{M+2}}^+|0>$, $\ldots$ , then a tensor product of the result
with $\phi_{b_{N-1}}^+|0>$, altogether $(N-M-2)$ times. Reproducing the considerations from the point (i), especially the formulae (\ref{pro}),(\ref{pr1})
we can see that this product contains a tensor with the index
structure described by $[1^{N-1}]$:
\be
   [1^M] \otimes[1] \otimes \ldots \otimes[1]=
[1^{N-1}]\oplus\ldots\ .\label{prN}
\ee
In terms of irreducible representations of $S_N$, (\ref{prN}) can be rewritten as:
$ (A)\times (N-1,1)\times\ldots\times(N-1,1)  =(B)\oplus\ldots$
where $(B)$ denotes the Young tableaux ,corresponding to the state
$\phi_{1}^+ \ldots\phi_{N-1}^+|0>$, and (A) is defined in (ii).

Let $(A)$ not contain the tableau $(N-(M+1),1^{M+1})$. As mentioned in
point (iii), the interior product of every Young tableau with $(N-1,1)$
contains only the Young tableaux differing from the initial one by no more
than the position of one cell. Therefore, the interior product of every
Young tableau $(N-M-2)$ times with $(N-1,1) $contains only the Young tableaux
differing from the initial one by no more than the position of $N-M-2$
cells. The rest three tableaux
 in (iii) differ from  $(1^N)$ by more than
$N-M-2$ cells because they all have less than $M+2$ cells in the first
column. So, (\ref{prN}) is not satisfied, unless $(A)$ contains the tableau
$(N-(M+1),1^{M+1})$.

(vi) The dimension of the space of states $\phi_{b_1}^+ \ldots\phi_{b_{M+1}}^+|0>$ is
equal to $C^{M+1}_{N-1}$, because it is a dimension of a subspace with a fermionic number equal to $M+1$, when the total fermionic number is equal to $N-1$.

The dimension of the Young tableau $(N-(M+1),1^{M+1})$ (and hence the dimension of the corresponding irreducible representation of $S_N$) is also equal to $C^{M+1}_{N-1}$, because \cite{Hammer} the dimension is the number of different ways of placing the integer numbers from $1$ to $N$ consecutively into the $N$ cells
of  the tableau, so that the number of the occupied cells not increase with the number of the line, and, placing each number into the line, we place it as close to its left end as possible.

At first, we place $1$ into the corner cell. Then, the distribution of the numbers in the cells  is determined unambiguously  by deciding what numbers we place into the $N-M-2$ lateral cells. It can be done in
$ C^{N-M-2}_{N-1}= C^{M+1}_{N-1}$ ways.

So, $(A)$ contains the tableau $(N-(M+1),1^{M+1})$ and nothing more. Therefore,
 the state $\phi_{b_1}^+ \ldots\phi_{b_{M+1}}^+|0>$ corresponds to an irreducible representation of $S_N$ with this Young
 tableau .

The proof is now completed.

\section*{\normalsize\bf Acknowledgements}

This work was made possible by support provided by Grant of
Russian Foundation of
Basic Researches (grant  99-01-00736).

\end{document}